\documentclass[%
 aip,
 amsmath,amssymb,
 reprint,%
]{revtex4-1}

\usepackage{graphicx}% Include figure files
\usepackage{dcolumn}% Align table columns on decimal point
\usepackage{bm}% bold math
\usepackage[utf8]{inputenc}
\usepackage[T1]{fontenc}
\usepackage{mathptmx}

\draft % marks overfull lines with a black rule on the right

\begin{document}

%\preprint{}

\title{Optical and microwave metrology at the 10$^{-18}$ level with an Er/Yb:glass frequency comb}

%Precision metrology made easy: a guide to the Er/Yb:glass optical frequency comb



\author{N. V. Nardelli}
 \email[Authors to whom correspondence should be addressed: ]{nicholas.nardelli@nist.gov and tara.fortier@nist.gov.}
  \affiliation{ 
National Institute of Standards \& Technology, 325 Broadway, Boulder, CO 80305, USA%\\This line break forced with \textbackslash\textbackslash
}%
\affiliation{%
University of Colorado, Boulder, Colorado 80309, USA%\\This line break forced% with \\
}%

\author{H. Leopardi}
\affiliation{ 
National Institute of Standards \& Technology, 325 Broadway, Boulder, CO 80305, USA%\\This line break forced with \textbackslash\textbackslash
}
\affiliation{%
University of Colorado, Boulder, Colorado 80309, USA%\\This line break forced% with \\
}%

\author{T. R. Schibli}
\affiliation{%
University of Colorado, Boulder, Colorado 80309, USA%\\This line break forced% with \\
}%

\author{T. M. Fortier}%
 \email[Authors to whom correspondence should be addressed: ]{nicholas.nardelli@nist.gov and tara.fortier@nist.gov.}
\affiliation{ 
National Institute of Standards \& Technology, 325 Broadway, Boulder, CO 80305, USA%\\This line break forced with \textbackslash\textbackslash
}%


\date{\today}

% approximately 250 words
\begin{abstract}
Optical frequency combs are an essential tool for precision metrology experiments ranging in application from remote spectroscopic sensing of trace gases to the characterization and comparison of optical atomic clocks for precision time-keeping and searches for physics beyond the standard model.  Here we describe the architecture and fully characterize a telecom-band, self-modelocking frequency comb based on a free-space laser with an Er/Yb co-doped glass gain medium. The laser provides a robust and cost-effective alternative to Er:fiber laser based frequency combs, while offering stability and noise performance similar to Ti:sapphire laser systems. Finally, we demonstrate the Er/Yb:glass frequency comb's utility in high-stability frequency synthesis using two ultra-stable optical references at 1157 nm and 1070 nm and in low-noise photonic microwave generation by dividing these references to the microwave domain.
\end{abstract}


\maketitle %\maketitle must follow title, authors, abstract and \pacs

%\tableofcontents

%%%%%%%%%%%%%%%%%%%%%%%%%%%%%%%%%%
%%%%%%%%%%%%%%%%%%%%%%%%%%%%%%%%%%
%%%%%%%%%%%%%%%%%%%%%%%%%%%%%%%%%%
\section{Introduction}

Just a few years after the first demonstration of the laser in 1960 \cite{Maiman1960}, 1963 brought the very first modelocked laser \cite{Hargrove1964}, which used an intra-cavity acousto-optic modulator to dynamically affect the cavity loss. The result was a highly regular train of pulses, which was the seed for numerous technologies, such as laser communication enabling trans-oceanic fiber optics and the modern internet. Pulse formation via shared phase coherence of laser spectral modes, or modelocking, yielded shorter pulses and higher pulse energies, permitting researchers new tools for both applied and fundamental research. High peak powers benefited applications such as soft-tissue ablation used in laser eye surgery \cite{Stern1989}, inertial confinement nuclear fusion \cite{Andiel2002}, which offers an alternative to magnetic confinement fusion, and two-photon absorption spectroscopy \cite{Denk1990}, which relies on highly efficient nonlinear interactions facilitated by the high optical intensity seen in pulses, just to name a few. Ultra-short pulses, with durations of femtoseconds or attoseconds, made possible ultra-precise measurements of time and distance, and are increasingly utilized in time-resolved measurements such as pump-probe experiments \cite{Lytle1985} that yield information about previously inaccessible ultra-fast processes. In fact, the modelocked laser is responsible for the shortest human-made event at 43 attoseconds \cite{Gaumnitz2017}.

Optical frequency combs (OFCs) based on modelocked lasers take advantage of the extreme precision with which the optical field can be controlled, and the nonlinear wavelength conversion made possible with highly energetic optical pulses, to enable high precision synthesis and control of the modelocked laser spectrum. Due to strong nonlinear optical interactions within a laser cavity \cite{Spence1991,Hofer1992,Keller1996}, frequency modes are forced to oscillate with a common phase and at rigid frequency intervals dictated by the length of the cavity. This phase-coherence creates a fixed frequency and phase relationship between all resonant optical frequency modes that extends to microwave signals generated via photodetection of the laser pulse train. This concept of optical-to-microwave synthesis makes possible the derivation of microwave timing signals from optical atomic clocks and the characterization of optical clocks against the current definition of the SI second, based on a microwave transition in the $^{133}$Cs atom \cite{Essen1955}. Additionally, ultra-low phase noise microwave signals have been demonstrated that take advantage of the coherent division of OFCs to divide the frequency and noise of narrow-linewidth cavity-stabilized optical references used to probe optical clock transitions \cite{Zhang2010,Haboucha2011,Fortier2013,Xie2017,Kalubovilage2020,Nakamura2020}. These room-temperature systems surpass the best cryogenic microwave systems and may enhance some applications such as low-noise Doppler radar by improving sensitivity to low-velocity and low-cross section objects \cite{Ye2000, Fortier2011, Xie2017}.

Because the optical spectrum of OFCs can span up to and beyond an optical octave of bandwidth, the optical spectrum of the modelocked laser is a coherent source of 100,000s to millions of equally spaced optical modes. The latter characteristic can enable the comparison of optical atomic clocks whose transition frequencies span 100s of terahertz, and with resolutions beyond the current limit imposed by the definition of the Hertz.

The optical-to-optical synthesis of high-accuracy and high-stability optical frequencies derived from atomic clocks via OFCs has helped to support a number of scientific and technological applications, including searches for Bosonic dark matter that are conducted by comparing optical clocks of different species \cite{Rosenband2008,Safronova2018}. Additionally, optically derived microwave signals could be used for the synchronization of radio telescopes in very long-baseline interferometry (VLBI) \cite{Clivati2017}, which is used to image cosmic radio sources. A high level of synchronization also benefits next-generation communication and navigation networks, such as GPS, enabling faster acquisition of location data with higher accuracy.

\begin{figure*}
\centering
\includegraphics[width=\textwidth]{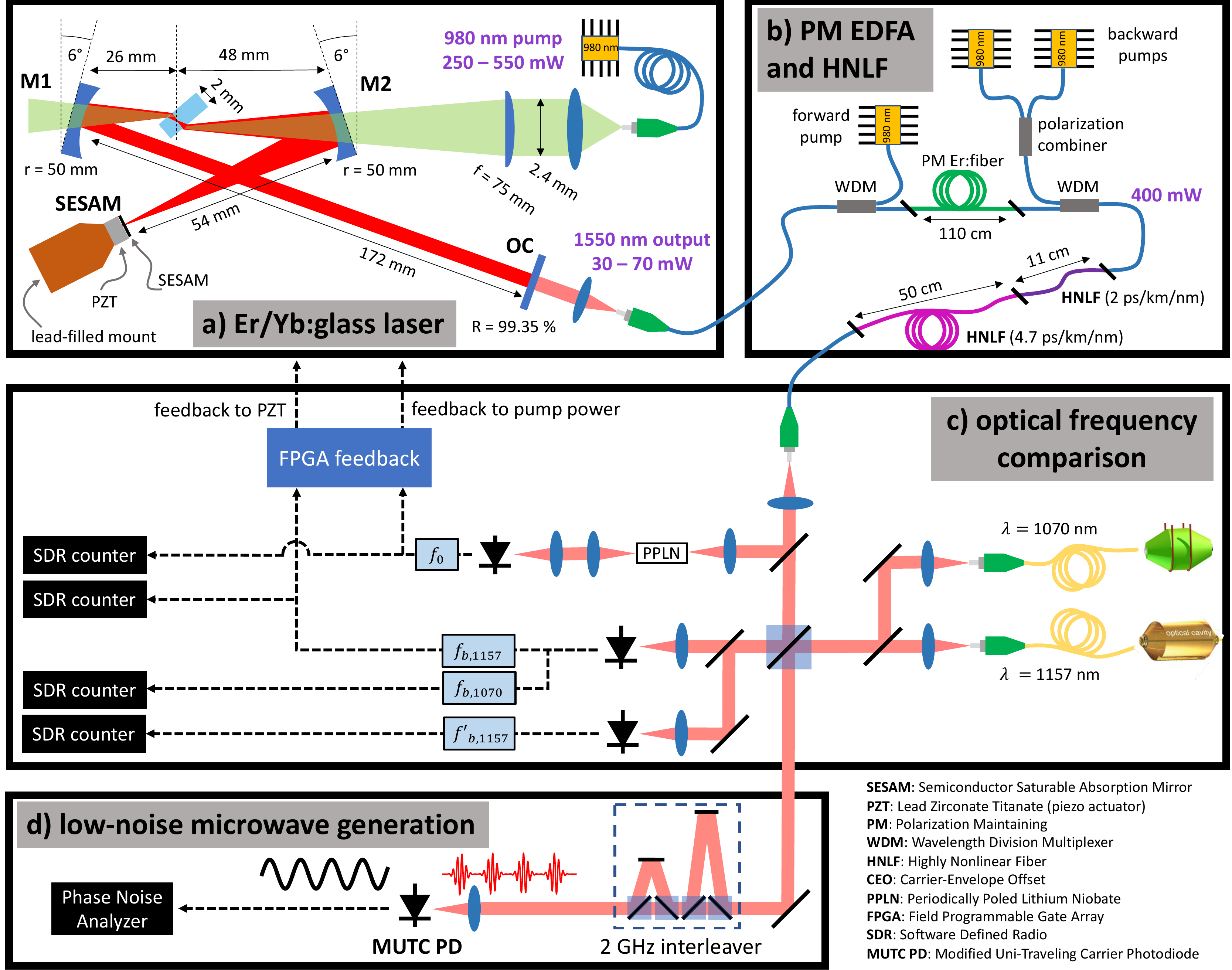}
\caption{Er/Yb:glass frequency comb experimental setup. a) The modelocked laser cavity architecture and geometry. b) External optical pulse amplification and supercontinuum generation using a PM-EDFA and PM-HNLF. c) Optical interferometers for referencing of the modelocked optical spectrum to optical references at 1157 nm and 1070 nm. A single optical reference is used for stabilization of the OFC repetition rate. The offset frequency $f_0$ is detected using a nonlinear $f - 2f$ interferometer. The detected optical beats are counted using zero-deadtime SDR counters. d) Low-noise optical-to-microwave generation. An interleaver to multiply $f_\text{rep}$ from 500 MHz to 2 GHz and an MUTC photodiode to convert the optical pulse train to the microwave domain for detection of the pulse repetition rate.}
\label{setup}
\end{figure*}

Optical frequency combs based on the Ti:Sapphire laser were used to realize the first OFCs \cite{Ye2000} and have been used in some of the highest performance metrology experiments with atomic clocks \cite{fortier201920,BACON2021}. However, the laser is prone to loss of modelocking as a result of high sensitivity to the ambient temperature drifts and cavity misalignment. It also requires a high-power, expensive visible pump laser (usually 532 nm) to excite emission in the gain bandwidth of the laser crystal. To address the latter drawback in robustness, the Er:fiber OFC \cite{Tauser2003} was developed as a more accessible system operating in the telecom band. OFCs based on Er:fiber lasers require almost no alignment after constituent fibers are spliced together, and employ off-the-shelf 980 nm pump diodes that have benefited from decades of development by the telecommunication industry. High loss in Er:fiber lasers, coupled with pump lasers with relatively high amplitude noise, results in resonant optical linewidths of 100 kHz to MHz. Also, due to the lengths of fiber that are required for intra-cavity gain and modelocking, fiber lasers generally do not support repetition frequencies higher than a few hundred MHz. For optical clock comparisons, this decreases the signal-to-noise ratio (SNR) of the optical beat between CW laser and OFC and makes phase-locking and detection more difficult. For low-noise microwave generation, lower repetition rates are associated with higher nonlinearity in photodetection.

In this work, we describe a more recently developed OFC based on a 500 MHz free-space laser with an Er/Yb co-doped phosphate glass gain medium \cite{Nardelli2022} that is designed to address the issues mentioned above. Due to the choice of gain medium, with an emission bandwidth near 1550 nm, the laser benefits from the same 980 nm pump diodes used in the Er:fiber laser. Modelocking is achieved via a semiconductor saturable absorber mirror (SESAM) \cite{Keller1990,Keller1996}, which permits robust self-starting and pulsed operation without precise cavity alignment. The laser also exhibits low intrinsic noise due, in part, to the low intra-cavity loss and the co-doping of Er and Yb that suppresses high-frequency pump intensity noise. We demonstrate the utility of this OFC in high-precision optical and microwave synthesis, which benefits next-generation optical atomic clock comparisons.

%%%%%%%%%%%%%%%%%%%%%%%%%%%%%%%%%%
%%%%%%%%%%%%%%%%%%%%%%%%%%%%%%%%%%
%%%%%%%%%%%%%%%%%%%%%%%%%%%%%%%%%%
\section{Er/Yb:glass OFC description and characterization}

In this section, we describe and characterize an Er/Yb:glass frequency comb, including the details of the modelocked laser oscillator, the amplification and supercontinuum generation, and the stabilization of the carrier-envelope offset frequency and the repetition frequency to an optical reference. Modelocked lasers based on Er/Yb:glass have been previously demonstrated with a repetition frequency of 1 GHz or less \cite{Lesko2020,Spuhler2005}, however, this laser has also been demonstrated with 10 GHz \cite{Resan2016} and 100 GHz \cite{Oehler2008} repetition frequencies. Additionally, the SESAM and Er/Yb:glass platform was demonstrated as a 1 GHz monolithic laser in bulk CaF$_2$ rather than free-space \cite{Shoji2016}.

%%%%%%%%%%%%%%%%%%%%%%%%%%%%%%%%%%
\subsection{Laser oscillator}

%%%%%%%%%%%%%%%%%%%%%%%%%%%%%%%%%%
\subsubsection{Er/Yb-doped phosphate glass}
An Erbium-doped gain medium constitutes a 3-level laser structure whereby 980 nm radiation drives electrons from the ground state ($^4$I$_{15/2}$) to the excited state ($^4$I$_{11/2}$). Energy is dissipated non-radiatively via a long-lived laser state ($^4$I$_{13/2}$) from where electrons are stimulated back to the ground state via 1.55 $\mu m$ photons. Direct pumping of Er$^{3+}$ exhibits low efficiency because of its weak absorption bands. Co-doping with Yb$^{3+}$ ions facilitates electron transfer and increased pumping efficiency of the gain medium. This is because Yb exhibits a very intense absorption band near 980 nm \cite{Gapontsev1982} and the Yb$^{3+}$ excited state ($^4$F$_{5/2}$) is resonant with the $^4$I$_{11/2}$ excited state in Er$^{3+}$.

Despite the low thermal damage threshold  of intense optical pumping, phosphate glass serves as a better gain material for 1.55 $\mu m$ lasers compared to other glasses and crystals. This is due to the presence of higher phonon energies that shorten the $^4$I$_{11/2}$Er$^{3+}$ lifetime to 2-3 $\mu s$, thus reducing the backwards transfer of energy from Er$^{3+}$ to Yb$^{3+}$ and increasing pumping efficiency \cite{Gapontsev1982}. 

%%%%%%%%%%%%%%%%%%%%%%%%%%%%%%%%%%
\subsubsection{Cavity geometry}
The Er/Yb:glass optical frequency comb employs a linear free-space laser cavity with a free spectral range of $f_\text{FSR} = f_\text{rep} \approx 500$ MHz (600 cm total optical path length), pictured in Figure \ref{setup} a). An advantage of the linear cavity design is that it permits tuning of the repetition frequency, via translation of the laser output coupler, by $>\pm$ 4 \% (from 480 MHz to 520 MHz) with minimal cavity misalignment. The laser glass medium is phosphate glass co-doped with Er$^{3+}$ and Yb$^{3+}$ (QX/Er), approximately 2 mm in length, with doping concentrations 0.8 \% wt. Er$^{3+}$ and 21 \% wt. Yb$^{3+}$, and an upper state lifetime of 7.9 ms. The relatively long upper state lifetime permits continuous pumping due to the large energy capacity of the gain medium, with negligible laser dynamics as no single pulse has enough energy to deplete the stored energy. We achieve a modelocked output power around 65 mW using a 0.65 \% output coupler centered at 1550 nm and a single 980 nm laser diode providing 450 mW of pump power. This supports transform limited optical pulses of 180 femtoseconds with pulse optical bandwidths of approximately 14 nm. Unfortunately, due to the low heat-handling capacity of phosphate glass, pumping the laser at powers higher than 600 mW introduces the risk of thermal fractures and cavity misalignment induced via thermal lensing and stress-induced birefringence.

\begin{figure}
\centering
\includegraphics[width=8.5cm]{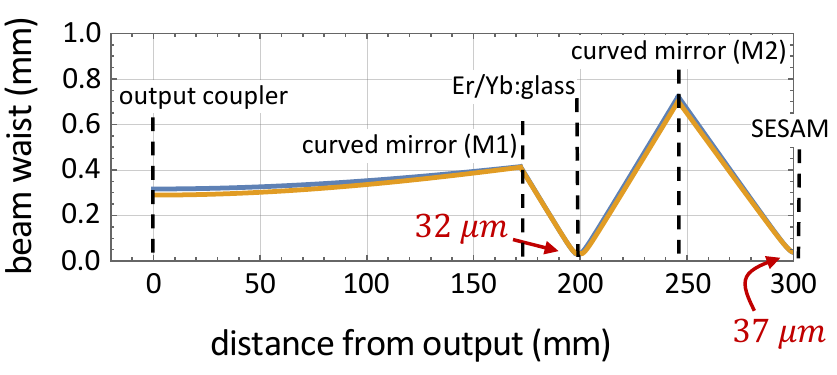}
\caption{Beam waist vs position in the cavity. The sagittal beam axis is shown in blue and tangential beam axis in orange. Laser mirror, Er/Yb:glass gain medium and SESAM positions are indicated by dashed black lines.}
\label{BeamWaist}
\end{figure}

We choose the cavity geometry to be asymmetric so that it yields two focuses (beam waist $\approx 35$ $\mu m$): one inside of the glass gain medium to support sufficient population inversion and nonlinearity, and a second at the surface of a SESAM, for passive modelocking. Figure \ref{BeamWaist} depicts how the beam waist varies over the cavity length, for both tangential and saggital dimensions, generated with an ABCD matrix calculation \cite{Kogelnik1966}. As seen in Figure \ref{BeamWaist}, the 1550 nm output is a collimated beam with 1/e$^2$ diameter around 0.6 mm. As a consequence of the cavity asymmetry, the Er/Yb:glass is off-center from the two curved mirrors. The single-mode 980 nm light is focused into the laser cavity via curved mirror M2 using a 7.5 cm lens. Each curved mirror has a 1550 nm reflectivity $> 99.9 \%$, a 980 nm transmission $> 98 \%$ and a radius of curvature, $r = 50$ mm.  A 6 degree angle between the curved mirrors and the optical axis helps counteract astigmatism introduced by the Er/Yb:glass, which is held at Brewster's angle ($\approx 57$ degrees with respect to the optical axis) to minimize light loss due to Fresnel reflections off the glass surfaces. Astigmatism expands the mode size in the tangential plane (parallel to the page) while preserving the mode size in the sagittal plane (perpendicular to the page) \cite{Hanna1969}. The latter is important for preserving cavity stability along both beam axes. While not employed here, coma aberration may also be compensated by arranging the cavity so that it has a Z geometry instead of an X geometry \cite{Dunn1977}.

%%%%%%%%%%%%%%%%%%%%%%%%%%%%%%%%%%

\subsubsection{Modelocking characterization}

For self-starting modelocking we employ a SESAM designed at the University of Colorado Boulder and grown at NIST that is composed of a single Er-doped, low-temperature-grown InGaAs quantum well deposited on top of a Bragg stack with alternating GaAs and AlAs layers, similar to the device reported by Lee et al. \cite{Lee2015}. Defects are intentionally introduced by growing the InGaAs at low temperature (475 C), allowing for ultrafast electron-hole recombination, which can facilitate the formation of intracavity femtosecond pulses \cite{Haiml1999,Mangold2013}. Slower dynamics between the conduction and valance bands (several picoseconds) allow intensity perturbations to grow for optical pulses as they make multiple passes through the laser cavity, leading to self-starting pulsed operation. 

The SESAM has greater than 99 \% reflectivity for fluences greater than 100 $\mu$J/cm$^2$, which corresponds to about 4 W of optical power circulating in the laser cavity. Typically, we operate the laser at about 10 W of circulating power and a fluence of about 300 $\mu$J/cm$^2$ on the SESAM. Nonsaturable loss accounts for the lost power, which is $<$ 1 \%. We found that SESAMs with higher nonsaturable loss ($>$ 3 \%) did not support stable pulse formation with this cavity geometry due to the limited gain possible with the 2 mm Er/Yb:glass.

\begin{figure}
\centering
\includegraphics[width=8.5cm]{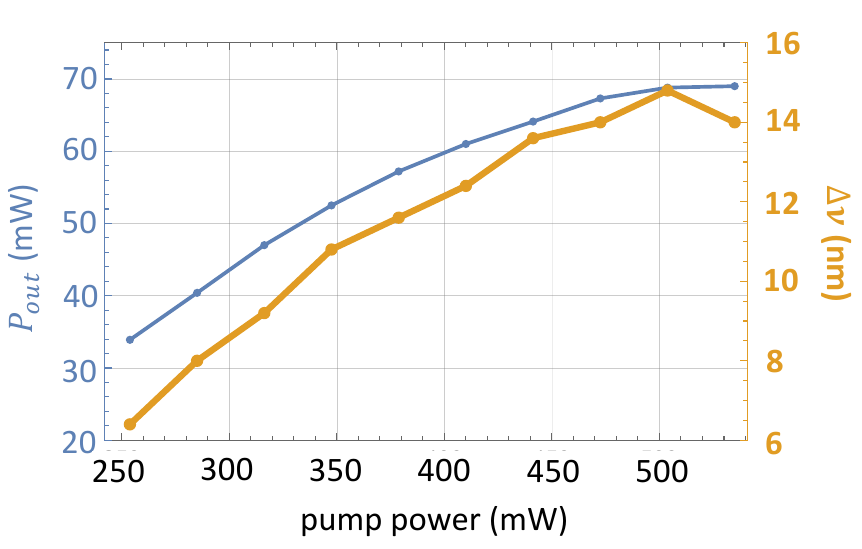}
\caption{Modelocked laser output power (thin trace, left axis) and pulse bandwidth (thick trace, right axis) versus 980 nm pump power.}
\label{CavityPowerWidth}
\end{figure}

We achieved stable pulse formation using 0.65 \% output coupling with as little as 250 mW of 980 nm pump power and an optical efficiency near 15 \%, decreasing at higher pump power (see Figure \ref{CavityPowerWidth}). The efficiency can be increased to 35 \% using an output coupler with 1.5 \% transmission, but at the sacrifice of modelocked optical bandwidth. This latter effect is due to the increased Er$^{3+}$ population inversion due to higher output coupling losses, which leads to a smaller gain bandwidth. Er/Yb:glass gain bandwidth is maximized for population inversions around 70 \% to 80 \% \cite{Spuhler2005}. From Figure \ref{CavityPowerWidth} we also observe that the laser can maintain stable modelocked operation for pump powers ranging from 250 mW to 550 mW, necessary for long-range control and stabilization of the carrier-envelope offset frequency (see Section on "OFC Phase Stabilization").

%%%%%%%%%%%%%%%%%%%%%%%%%%%%%%%%%%
\subsection{Amplification and supercontinuum generation}

The 500 MHz pulse train generated by the modelocked laser is coupled into polarization-maintaining (PM) fiber and amplified in a home-built PM Erbium-doped fiber amplifier (EDFA). The higher repetition rate of the Er/Yb:glass laser requires amplification to 400 mW to achieve the pulse energies required ($\approx$ 0.8 nJ) for supercontinuum generation in highly nonlinear fiber (HNLF). As a result, amplification requires three 980 nm pump diodes with a total power of about 2.5 W to pump a 110 cm length of Erbium-doped fiber, shown in Figure \ref{setup} b).

The amplified pulses then traverse $\approx$1 m of hybrid HNLF \cite{Korel2014}, which spreads the $\approx$14 nm input pulse across an optical octave of bandwidth from 1000 nm to 2200 nm. The HNLF consists of an 11 cm length of HNLF [4.7 ps/(km$\cdot$nm) dispersion] followed by a 50 cm length of HNLF [2 ps/(km$\cdot$nm) dispersion]. The high-dispersion short length of fiber serves to generate a broad dispersive wave that contains frequencies from about 1000 nm to 1200 nm, which covers the wavelengths of NIST's two optical atomic clock lasers at 1070 nm and 1157 nm \cite{Schioppo2016,Young1999} and a region around 1035 nm.  The low-dispersion longer length of fiber serves to spread optical power to longer wavelengths to access the region around 2070 nm. Light from the HNLF at 1035 nm and 2070 nm is used for self-referenced detection of the carrier-envelope offset frequency \cite{Telle1999,Reichert1999,Jones2000}.

\begin{figure}
\centering
\includegraphics[width=8.5cm]{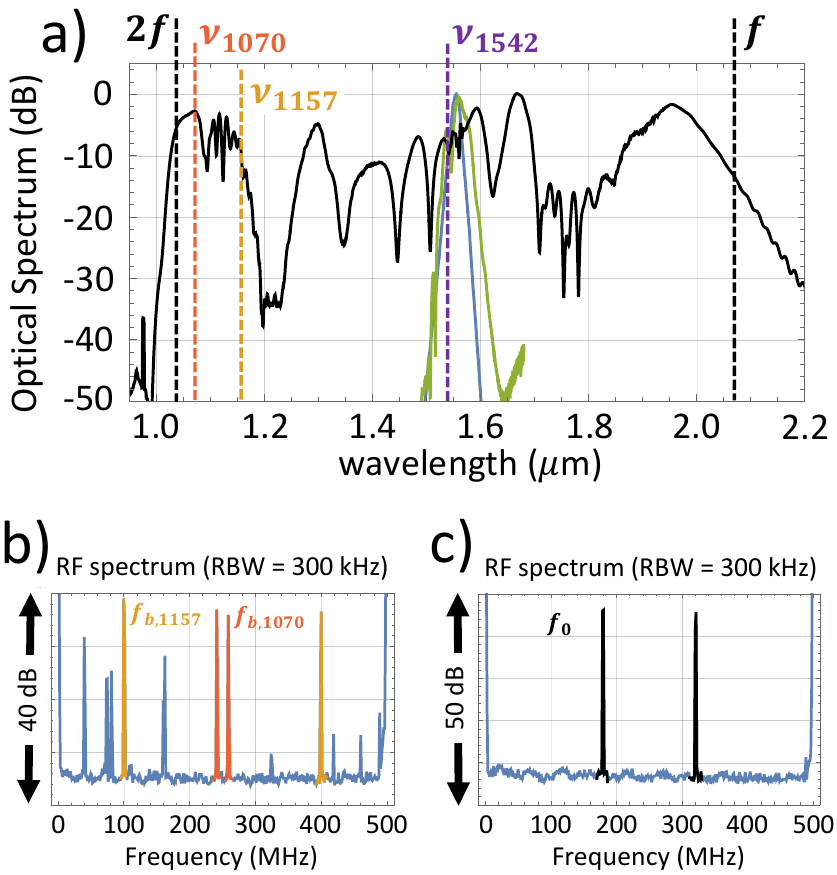}
\caption{a) Optical pulse spectrum as measured at different points in the OFC experimental setup: at the output of the modelocked laser (blue), and after amplification with the PM EDFA (green), after broadening with HNLF (black). Dashed lines indicate the frequencies of the optical references at 1157 nm (orange), 1070 nm (red) and 1542 nm (purple) and the $f$ and $2f$ frequencies used for carrier-envelope offset detection (black). In the bottom, RF spectra of the heterodyne beat signals between the comb and optical references, b), and of the $f - 2f$ interferometer, c). Spectra are measured with a resolution bandwidth of 300 kHz.}
\label{opticalSpectrum}
\end{figure}

Figure \ref{opticalSpectrum} a) depicts the optical spectrum, with dotted lines indicating both the clock frequencies and the f-2f frequencies. A line at 1542 nm is included, which indicates the frequency of the $^{87}$Sr lattice clock's cryogenic Silcon cavity stabilized laser at JILA \cite{Matei2017}. This supercontinuum spectrum can be shaped and tuned to a small degree by controlling the power and dispersion of the pulses launched into the HNLF. Higher pulse energies push optical power out to the $f$ and $2f$ frequencies at the expense of power at the 1157 nm clock laser frequency. We are able to further shape the spectrum by adding short lengths of fiber between the EDFA and the HNLF to pre-chirp and temporally compress pulses as they traverse the HNLF. The latter enables a longer interaction length and more efficient spectral broadening.

The optical heterodyne beats between comb and optical clock light and the carrier-envelope offset frequency are shown in Figure \ref{opticalSpectrum} b) and c), respectively. Typically, an SNR of greater than 30 dB in a 300 kHz bandwidth ensures that a beat signal is adequately counted and stabilized with minimal cycle/phase slips. Our single-branch optical amplifier and HNLF architecture is sufficient to meet the latter criteria. However, one can potentially improve SNR by shaping the optical spectrum using a nonlinear pulse propagation solver to optimize the HNLF length, or by directly engineering the HNLF dispersion with patterned Bragg gratings \cite{Westbrook2004}, or engineered nonlinear waveguides to enhance specific regions of the spectrum \cite{Carlson2017}.

% \subsubsection{Amplitude noise}
Additionally, amplitude noise is important to consider given the potential impact of amplitude-to-phase conversion in nonlinear optical and electronic components. For instance, when using a frequency comb to synthesize low-noise microwave signals, nonlinearity in the photodetector converts amplitude noise on the optical pulse train to phase noise on the detected microwave carriers \cite{Taylor2011,Fortier2013}. 

\begin{figure}
\centering
\includegraphics[width=8.5cm]{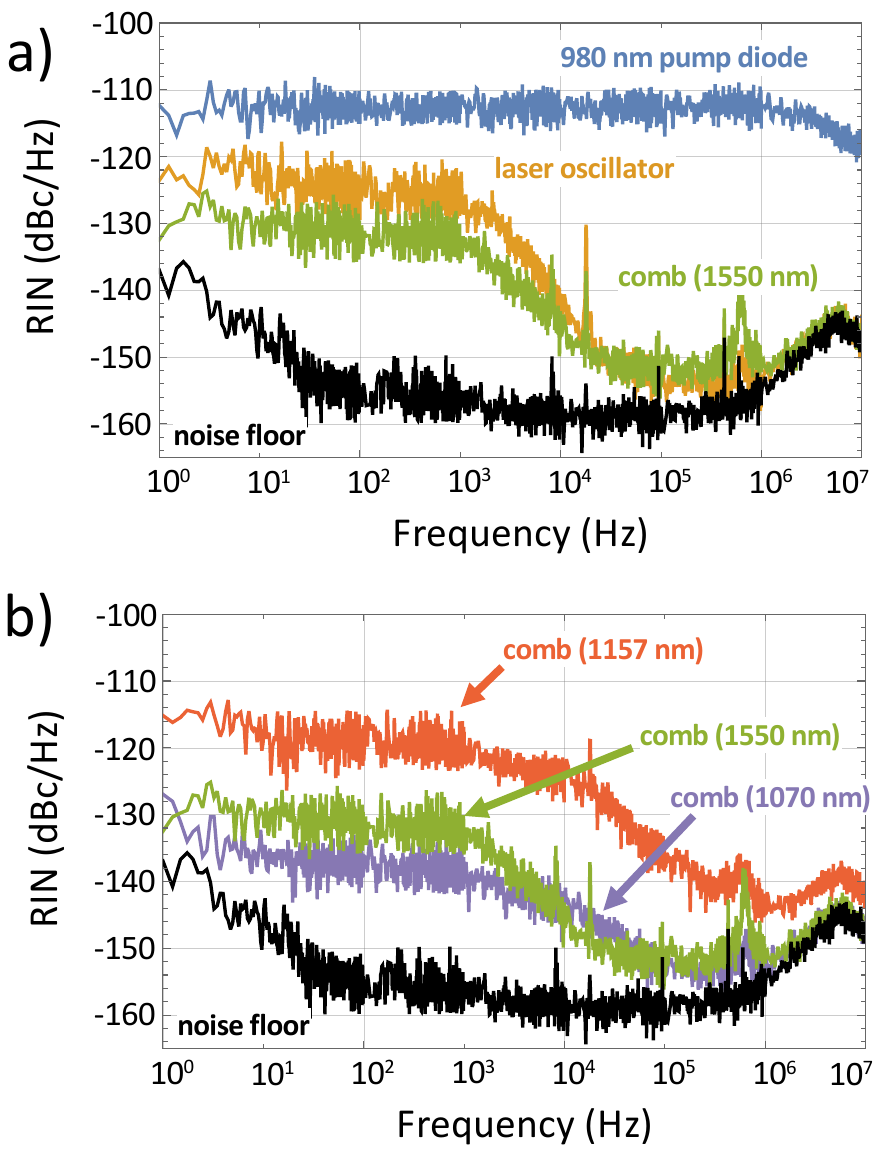}
\caption{a) Relative intensity noise as measured at different points in the OFC experimental setup: the 980 nm pump laser (blue), the modelocked laser output (orange), the pulse train at the output of the HNLF, filtered around 1550 nm (green). b) Relative intensity noise measured at the output of the HNLF, filtered around 1157 nm, 1550 nm and 1070 nm. The measurement noise floor in a) and b) is shown in black.}
\label{RIN}
\end{figure}

Figure \ref{RIN} a) depicts the suppression of relative intensity noise (RIN) at different points in the optical frequency comb, from the 980 nm pump, to the modelocked laser and to the output of the HNLF. As seen in Figure \ref{RIN} a), RIN on the 1550 nm pulse train output from the laser cavity appears to be suppressed compared to the RIN of the 980 nm pump laser. The laser cavity RIN is further suppressed by operating the EDFA in saturation, which clamps the output power. From Figure \ref{RIN} b) it is clear that RIN is not equal among the different frequency bands of interest, with the 1157 nm band displaying much more intensity noise than the other bands. This is likely due to the fast roll-off in wavelength-dependent intensity around 1157 nm as seen in Figure \ref{opticalSpectrum} a). Also seen in these plots is a large feature around 10 MHz, which is due to the transimpedance amplifier of the photodetector circuit.

The Er/Yb co-doping renders this laser relatively insensitive to pump laser intensity noise since the Yb ions primarily absorb the pump photons and the Er ions primarily emit photons \cite{Taccheo1998}. Three-level lasers, such as the Er:fiber laser, are directly pumped and so are more sensitive to pump intensity noise. The roll-off around 1 kHz is a result of the Yb ion upper-state lifetime and energy transfer rate to Er ions, which acts as a low-pass filter for intensity noise \cite{Lee2014}. The higher-frequency RIN originates instead from cavity losses due to acoustic and mechanical environmental noise.

%%%%%%%%%%%%%%%%%%%%%%%%%%%%%%%%%%
\subsection{OFC Phase Stabilization}

Pulse formation in the modelocked laser forces all resonant longitudinal cavity modes to share the same phase. This shared phase coherence allows the optical modes to be connected to one another via a simple and fixed frequency relationship, known as the comb equation. As a result, the $N^{\text{th}}$ mode of the comb is defined by the mode number and the pulse repetition frequency and the laser offset frequency, such that 

\begin{equation}
    \nu_N = N f_\text{rep} + f_0. 
\end{equation}

Note that we use the symbols $f$ and $\nu$ to denote RF and optical frequencies, respectively. The latter equation has a multitude of important consequences regarding OFC operation and measurement capabilities \cite{fortier201920}. The definition of OFC optical modes with RF frequencies yields a direct optical-to-microwave frequency link, which permits SI traceability of all OFC optical modes, multiplication of microwave frequencies to the optical domain, and division of optical frequencies to the microwave domain. Additionally, shared phase coherence across the OFC optical modes also permits the comparison of optical references separated by more than 100 THz in frequency. The latter is possible since the regularly spaced optical modes of the comb can be used to measure the relative frequency differences, as well as the relative phase deviations between multiple optical references. Stabilization of the modelocked laser is achieved by detecting and then phase-locking $f_0$ and $\nu_N$ to high stability frequency references.

\textbf{Repetition rate detection and stabilization}  The $M^{\text{th}}$ mode of the laser repetition rate is stabilized to an optical reference by detecting and controlling the difference frequency (i.e., beat frequency), $f_{b,1157} = \nu_{1157} - (M f_\text{rep} + f_0) < 500$ MHz, between the nearest neighbor comb mode, $M$, and a cavity-stabilized laser at 1157 nm, $\nu_{1157}$. The 1157 nm laser serves as the local oscillator to the $^{171}$Yb optical lattice clock via doubling to the clock transition frequency near 578 nm. 

As shown in Figure \ref{setup} c), light output from the HNLF is split across various wavelength ranges and sent to optical interferometers. To minimize noise due to uncommon fiber-optic paths, light from the HNLF is coupled to free space. To a lower extent than fiber paths, uncompensated free-space paths contribute added instability to the detected heterodyne beat signals between the comb and optical references as a result of refractive index changes due to air currents, as well as from length changes due to expansion of the optical breadboard. To mitigate the effects of the additive instability, comb light that is transmitted through the dichroic mirror is interfered on a single beam splitter with the combined light from two ultra-stable optical references at 1157 nm and 1070 nm.  In this configuration, the total uncompensated free-space path length is 20 cm. 

Optical filters are used in the detection of all optical beat signals to reduce the light noise in photodetection from non-contributing optical comb modes. The 1157 nm beat signal is filtered and amplified before being mixed with a 100 MHz synthesized frequency derived from a H-Maser. The filtered error signal is fed back to a piezo-electric actuated mirror to compensate frequency deviations and phase noise due to instabilities in laser cavity length. To maximize the actuator bandwidth, a miniature piezo stack is affixed to a lead-filled copper mount, whose asymmetric geometry is chosen to damp acoustic resonances \cite{Briles2010}. Due to the light-weight SESAM ($\approx$ 3 mm $\times$ 3 mm $\times$ 1 mm), and sturdy PZT mounting structure, we observe a closed loop $f_{b,1157}$ servo bandwidth of >150 kHz. Referencing of the optical beat signal to a H-maser reference and filtering of the subsequent error signals is achieved using a PID algorithm in a field programmable gate array (FPGA) \cite{Pomponio2020}. Because the algorithm is implemented digitally, we are able to control and optimize the proportional (P), integral (I) and differential (D) gains and corner frequencies via a computer interface. The design of high bandwidth feedback actuators helps to suppress high-frequency comb noise and support higher performance frequency synthesis.

\textbf{Offset frequency detection and stabilization} We detect the offset frequency using a nonlinear self-referencing technique that compares the frequency-doubled low-frequency portion of the octave-spanning spectrum generated from the HNLF to the fundamental high-frequency portion of the spectrum. The latter technique is necessary since $f_0$ is only manifest on the optical carrier phase, which is inaccessible via direct photodetection. 
 
Using the self-referencing technique, the offset frequency is detected as $f_0 = 2 \nu_N - \nu_{2N} = 2 (N f_\text{rep} + f_0) - (2N f_\text{rep} + f_0)$. The optical spectrum around 1035 nm and 2070 nm is split off with a dichroic mirror and focused into a periodically poled lithium niobate (PPLN) waveguide where light at 2070 nm is frequency-doubled to 1035 nm. The interference between the fundamental 1035 nm light and doubled 2070 nm light is detected as $f_0$. Control and stabilization of the detected $f_0$ beat signal is achieved via comparison to a H-maser synthesized RF frequency. The resulting error signal is sent through a second FPGA PID loop filter and fed back to the Er/Yb:glass laser 980 nm pump current. Changes in the 980 nm pump power modify the gain of the Er/Yb:glass laser, which permits tuning of $f_0$, whereby $\approx$ 10 MHz of tuning is achieved by changing the 980 nm pump power by 50 mW.

%%%%%%%%%%%%%%%%%%%%%%%%%%%%%%%%%%
\subsection{OFC stability and noise characterization}

Once the modelocked laser offset frequency and repetition rate are stabilized, the resultant frequency comb represents a versatile tool for high precision optical and microwave frequency metrology. Assessing its stability and accuracy at different timescales is necessary to ensure it can support the most stringent of measurements. Also, because the relevant timescales differ by application, below we evaluate the stabilized Er/Yb:glass laser performance for timescales < 1 s by measuring the single-sideband phase noise (SSPN), and for time scales > 1 s by assessing the frequency stability of the time record of frequency counted data. The latter metrics can be used to evaluate both the "in-loop" and "out-of-loop" performance. The in-loop noise indicates how well the laser phase-locked loop is able to suppress the noise of the modelocked laser when it is stabilized to a low-noise reference. The out-of-loop measurement assesses the performance of the full system, including the OFC synthesis, uncompensated noise by the feedback loop, and the performance of the frequency reference.

%%%%%%%%%%%%%%%%%%%%%%%%%%%%%%%%%%
\subsubsection{OFC performance at short timescales: phase noise}

\begin{figure}
\centering
\includegraphics[width=8.5cm]{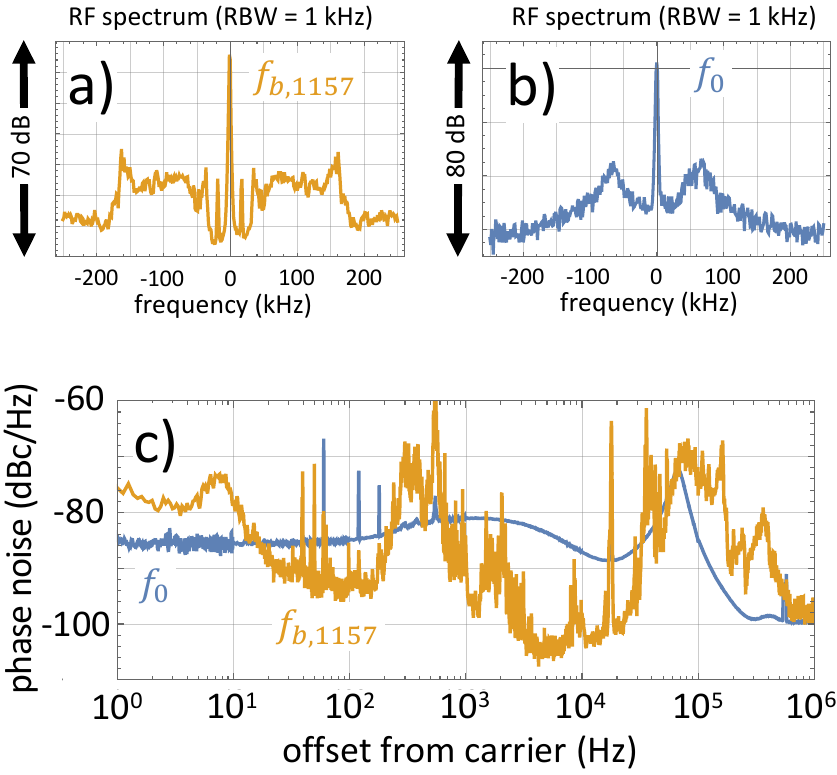}
\caption{RF spectra of the a) stabilized optical beat between the comb and the optical reference (left), $f_{b,1157}$ and b) carrier-envelope offset frequency, $f_0$. The RF spectra are measured with a resolution bandwidth of 1 kHz. c) Single-sideband phase noise spectra of the in-loop $f_{b,1157}$ beat signal (orange) and in-loop $f_0$ beat signal (blue). The latter yielded a RMS integrated phase noise (from 1 Hz to 1 MHz) of the 188 mrads and 80 mrads, respectively. }
\label{beats}
\end{figure}

Figure \ref{beats} a) and b) show the RF spectrum (1 kHz bandwidth) of the in-loop stabilized $f_0$ and $f_{b,1157}$ beat signals. The resolution-limited coherent carriers at 102 MHz and 179 MHz indicate that the $f_{b,1157}$ and $f_0$ are phase locked with an additive noise linewidth less than 1 kHz. 

Higher resolution frequency and phase noise measurements can be achieved by demodulating the microwave carriers. Figure \ref{beats} c) shows the single-sideband phase noise spectrum as measured using an RF network analyzer. The SSPN represents the power spectral density of the noise sidebands to one side of the RF spectrum. As mentioned previously, the phase noise power spectral density of the in-loop signal measures the additive noise of the Er/Yb:glass laser when it is stabilized to a low-noise frequency reference. As seen in Figure \ref{beats} b), we observe a servo bandwidth for the $f_0$ feedback loop near 60 kHz. The bandwidth of the $f_0$ feedback loop was increased beyond the 10 kHz upper state lifetime limit of the Er/Yb:glass gain medium by adjusting the differential gain and corner frequency of the FPGA-based PID filter. We also observe (Figure \ref{beats} a)) a 160 kHz $f_\text{rep}$ feedback bandwidth of the piezo-actuated cavity mirror on the $f_{b,1157}$ SSPN. While control and stabilization of $f_\text{rep}$ via the cavity length results in minimal frequency deviations on $f_0$, we observe that stabilization of $f_0$, via modulation of the 980 nm pump power pump, writes noise onto the laser repetition as observed by the shared 60 kHz sidebands on the $f_0$ and $f_{b,1157}$ beat signals and SSPN. Additionally, we observe a large, broad noise feature on the $f_{b,1157}$ SSPN between 100 Hz and 10 kHz, which is a result of unsuppressed cavity length noise due to acoustic and mechanical vibration of the modelocked laser.

Using the SSPN we calculate the additive integrated phase noise, pulse-to-pulse timing jitter and noise effective linewidth. We estimate that the root-mean-squared (RMS) integrated phase noise from 1 Hz to 1 MHz is 80 mrads and 188 mrads for $f_0$ and $f_{b,1157}$, respectively. This accounts for respective RMS timing jitters of 44 attoseconds (on 290 THz) and 115 attoseconds (on 259 THz). From the integrated phase noise, we can also determine that the phase-locked linewidths are less than 1 Hz, limited by resolution of the phase noise analyzer. The linewidth of $f_{b,1157}$ is ultimately limited by the high-stability 1157 nm reference laser (around 10 mHz) and the linewidth of $f_0$ is ultimately limited by the H-maser reference (around 1 $\mu$Hz).

%%%%%%%%%%%%%%%%%%%%%%%%%%%%%%%%%%
\subsubsection{OFC performance at longer timescales: frequency counting}

Optical frequency comparisons of like and unlike species of atomic clock are necessary in the characterization and development of optical atomic clock technology. While like clocks can be compared directly to one another, inter-species clock comparisons require an OFC to bridge the vast differences that separate transition frequencies. Aside from the application to the development of time and frequency references, the relative comparison of clock transition frequencies permits clocks to be high-sensitivity sensors of space-time variation  for relativistic geodesy \cite{McGrew2018,Grotti2018,Delva2019} and for searches for physics beyond the standard model \cite{Rosenband2008,Safronova2018}.

For clock comparisons of unlike species, the additive noise of the optical frequency comb is required to be lower than the clocks being compared. Currently, the best optical clocks have demonstrated fractional frequency accuracies around $10^{-18}$ and fractional instabilities around $10^{-16}$ after 1 second of averaging \cite{Huntemann2016,Huang2022,Brewer2019,McGrew2018,Bothwell2019}.  Below, we demonstrate that the Er/Yb:glass comb is able to achieve an additive measurement instability at least one order of magnitude lower than the current best optical atomic clocks comparison.

As mentioned previously, frequency counting can be used to assess the stability of the OFC at timescales greater than 1 s. For the results presented here, signals from the OFC were measured using home-built, zero-dead time, frequency counters employing software-defined radio (SDR), which offers a reconfigurable platform for digital signal processing. Our SDR counter is configured using two channels, which allows for cancellation of noise introduced by the SDR's numerically controlled oscillator, thus permitting zero-dead time frequency counting with 12-digits of frequency resolution and additive frequency offsets of $10^{-13}$ on a counted 10 MHz signal. Details regarding our SDR frequency counters can be found in Ref. \cite{Sherman2016}. Zero-dead time counting permits phase correlations to be maintained between frequency measurements, which helps to supports accurate frequency statistics for signals limited by white phase noise. The latter is important for characterizing the signals from optical frequency combs, which are expected to exhibit phase coherence in optical-to-optical, and in optical-to-microwave synthesis.

To assess the long-term performance of the OFC, the sampled input signal to an SDR counter was averaged over 1 s to deduce its average frequency. Consecutive measurements produce a time record of frequencies, averaged over 1 s and spaced by 1 s. 
Using the latter time record of frequency measurements, we calculate the two-sample overlapping Allan deviation, which yields the fractional frequency stability, versus measurement time. For signals dominated by white frequency noise and white phase noise, the fractional frequency stability improves with longer averaging time, $\tau$, with respective slopes of $\tau^{-1/2}$ and $\tau^{-1}$. 

\begin{figure}
\centering
\includegraphics[width=8.5cm]{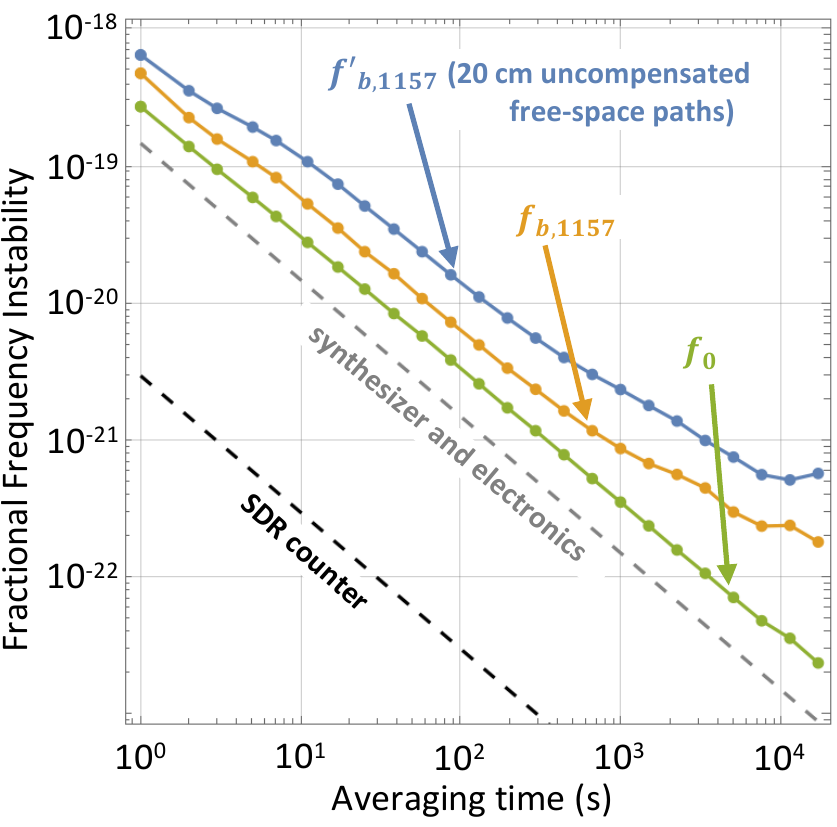}
\caption{Fractional frequency stability of the in-loop $f_{b,1157}$ (orange), $f'_{b,1157}$ (blue) and $f_0$ (green) beat signals as calculated via Allan deviation. The $f'_{b,1157}$ beat measures the noise contribution from 20 cm of out-of-loop optical free-space paths. Frequency noise from the H-maser referenced synthesizer and associated electronics to the optical beat signal (grey). Resolution limit contributed by the SDR frequency counter (black dashed line).}
\label{residualAdev}
\end{figure}

As seen in Figure \ref{residualAdev} the Allan deviation of the in-loop $f_0$ and $f_{b,1157}$ beat signals both follow slopes at, or close to, $\tau^{-1}$ for all averaging times, indicating shot or thermal noise as the limiting noise source. While in-loop signals mainly assess the quality of the laser feedback actuators, for OFCs with single-branch architectures \cite{Leopardi2017} they can provide relatively good estimates for the total residual, or additive noise, that the modelocked laser contributes in optical synthesis. This is possible if all optical fibers after the modelocked laser are within the stabilization loops and their noise is compensated via feedback to the laser.

\textbf{Er/Yb:glass laser OFC: sources of additive instability}
In this OFC, uncompensated free-space paths, and optical and electronic components are the main source of differential noise when detecting a common wavelength. Shown in Figure \ref{residualAdev} are the additive instabilities from the SDR, the RF synthesizer and microwave electronics (RF amplifiers, mixers and filters). We can evaluate the noise due to uncommon free-space optical paths in the Er/Yb:glass OFC system by detecting the $f_{b,1157}$ beat on two photodetectors. The in-loop $f_{b,1157}$ beat signal is used to stabilize the modelocked laser cavity length, and the second beat signal $f'_{b,1157}$ measures the 20-cm differential free-space paths, and noise added in photodetection between optical interferometers, seen in Figure \ref{setup} c). Figure \ref{residualAdev} shows the stability of the frequency detected on the in-loop detector (orange) and on the second detector (blue). From the latter measurement we observe that 20-cm of enclosed and uncommon free-space optical paths contribute a frequency instability near $7 \times 10^{-19}$ at 1 s. Noise resulting from changes in the index of refraction of air reduces the phase coherence, which alters the slope in the Allan deviation from $\tau^{-1}$ to a slope closer to $\tau^{-1/2}$.

%%%%%%%%%%%%%%%%%%%%%%%%%%%%%%%%%%%%%%%%%%
%%%%%%%%%%%%%%%%%%%%%%%%%%%%%%%%%%%%%%%%%%
%%%%%%%%%%%%%%%%%%%%%%%%%%%%%%%%%%%%%%%%%%
\section{Stability and accuracy in optical frequency synthesis}

At NIST, optical frequency combs are used to characterize and compare the $^{171}$Yb and $^{87}$Sr neutral lattice clocks \cite{McGrew2018,Bothwell2019} and the $^{27}$Al$^+$ quantum logic clock \cite{Brewer2019}. While the clocks operate on transition frequencies of 518 THz, 489 THz and 1.12 PHz, respectively, OFCs measure the clock local oscillator frequencies of 259 THz (1157 nm), 195 THz (1542 nm), and 280 THz (1070 nm). 

\begin{figure*}
\centering
\includegraphics[width=17cm]{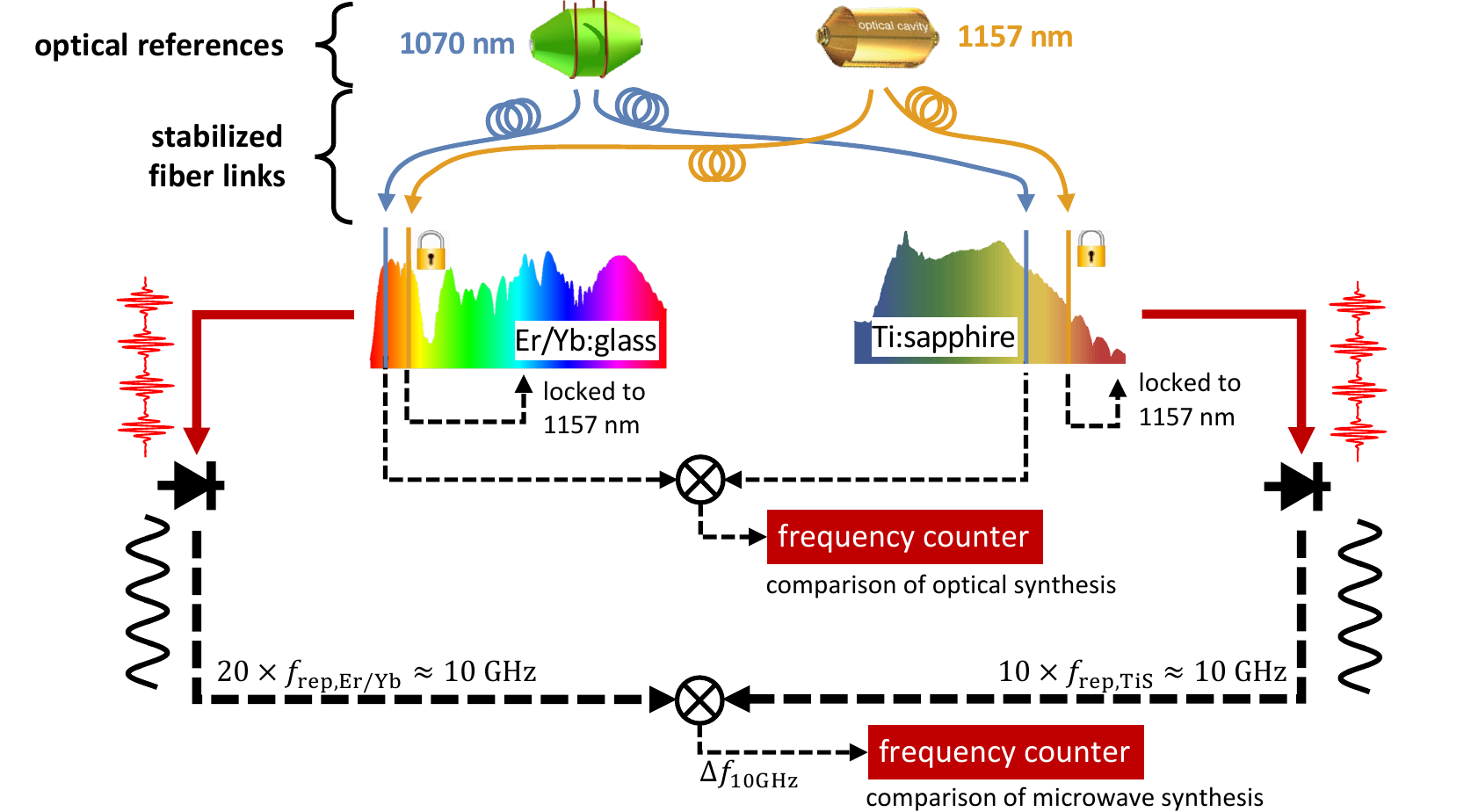}
\caption{Setup to characterize the Er/Yb:glass frequency comb in optical-to-optical and optical-to-microwave synthesis. Light is distributed from optical references at 1070 nm and 1157 nm over phase-stabilized fiber links to the Er/Yb:glass comb and a Ti:sapphire comb. Both combs are stabilized to the 1157 nm optical reference and light synthesized at 1070 nm from the two OFCs is compared using a common 1070 nm optical reference. Optically derived microwave at 10 GHz signals, generated via detection of the OFC optical pulse trains detecting the comb pulse trains on MUTC photodiodes, are also compared.}
\label{characterization-setup}
\end{figure*}

While optical synthesis via four wave mixing is expected to maintain the equidistance of the modelocked laser optical modes and preserve their phase coherence, tests to evaluate the noise in optical synthesis are necessary to assess the full performance of the OFC for optical atomic clock comparisons. Nonlinear techniques have been used to evaluate synthesis between OFC modes separated by one octave using a single comb \cite{Stalnaker2007,Leopardi2017,Martin2009,Stenger2002}. However, assessing synthesis errors over smaller bandwidths generally requires comparing the relative output of two OFCs. Below we evaluate the additive instability in optical synthesis of the Er/Yb:glass comb between 1070 nm and 1157 nm via comparison to a second frequency comb based on a Ti:sapphire modelocked laser \cite{Fortier2006}, see Figure \ref{characterization-setup}.

\begin{figure}
\centering
\includegraphics[width=8.5cm]{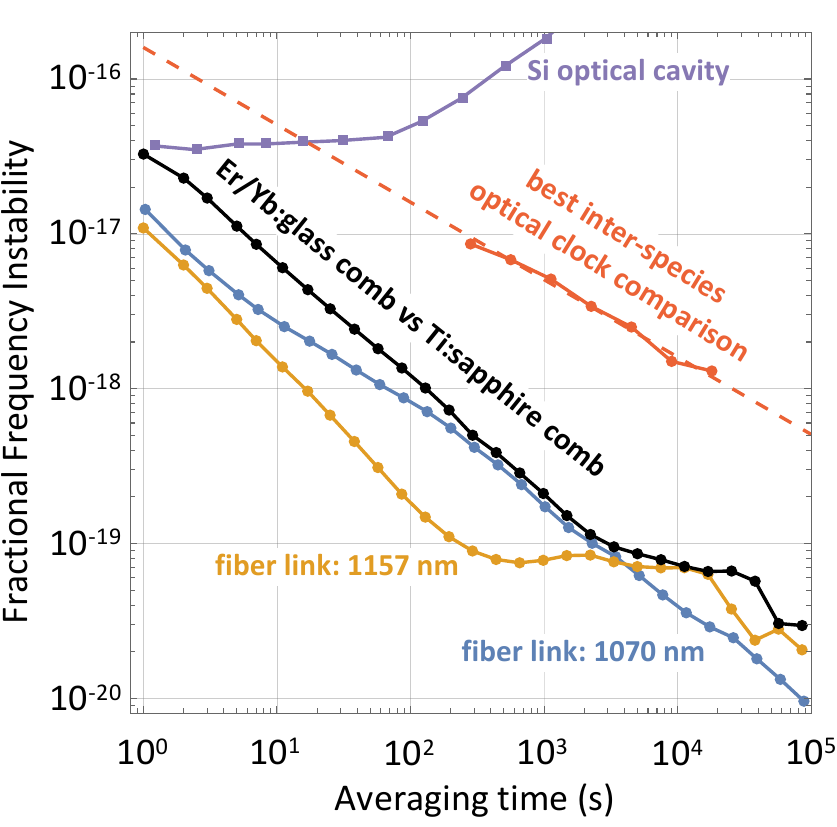}
\caption{Comparison of fractional frequency instabilities for various optical-to-optical comparisons. Comparison of the noise in optical-to-optical synthesis from 1157 nm to 1070 nm between the Ti:sapphire and Er/Yb:glass OFCs (black points). Optical beat note between two 50 m 1157 nm fiber links (orange trace). Optical beat note between two 100 m 1070 nm fiber links (blue trace). The current best inter-species optical clock comparison \cite{Kim2021} between the $^{27}$Al+ ion clock and $^{171}$Yb optical lattice clock using differential spectroscopy (red points) fit to a white frequency noise trend line ($1.6 \times 10^{-16} ~\tau^{-1/2}$), shown with red dashed line. The purple squares represent a high-stability, cryogenic Si optical reference cavity \cite{Matei2017}. }
\label{combAgreement}
\end{figure}

The reference cavities used to stabilize the 1157 nm and 1070 nm clock lasers are shown in Figure \ref{characterization-setup} and their light was distributed over independent phase-stabilized optical fiber links \cite{Ma1994}. In the measurement, both OFCs were phase-locked to the 1157 nm reference. Although the OFCs have different offset frequencies and repetition rates, stabilization of both comb spectra to the 1157 nm reference should faithfully, and identically, transfer the optical reference phase and frequency information to all OFC synthesized optical modes, including those near 1070 nm. Errors in synthesis can be measured by comparing the detected beat signals between the 1070 nm reference and the nearest mode for each comb. The latter comparison was achieved by filtering and amplifying each 1070 nm beat signal and deriving an RF signal at the difference frequency of the two beats using a double balanced mixer. The difference frequency was counted on an SDR frequency counter and the Allan deviation was calculated (black points in Figure \ref{combAgreement}). Because the optical references were free-running, we set the repetition frequency of the Ti:sapphire comb, $f_\text{rep,TiS}$, to be exactly twice that of the Er/Yb:glass comb, $f_\text{rep,Er/Yb}$, so that the scaling factor between 1157 nm and 1070 nm frequency modes for the two combs was identical. This allowed the frequency drift of the optical references to cancel in the stability measurement.

From the figure, the data represented by the black points exhibits a 1-s instability near $3 \times 10^{-17}$, averaging down to the $10^{-20}$ level, limited by the total measurement time. It is important to note that the observed stability represents the combined noise of the four optical fiber links, linking the optical references to the OFCs, as well as the noise in synthesis from two OFCs. As a result, the observed instability represents an upper limit to the residual noise in optical synthesis. More specifically, we found that the latter optical synthesis measurement was partly limited by the frequency stability of the Doppler-stabilized fiber links. The fiber link stabilities were assessed by measuring a beat signal between light from the two 1157 nm links and a separate beat signal between light from the two 1070 nm links. The Allan deviations for both sets of links are plotted in orange and blue, respectively, in Figure \ref{combAgreement}. These demonstrate that the Doppler-cancelled fiber links contributed instabilities near $10^{-17}$ at 1 second averaging times, averaging down to levels near $10^{-20}$ at longer times. 

Also shown for comparison in Figure \ref{combAgreement} is the current best stability achieved for an inter-species optical clock comparison \cite{Kim2021}, which yielded a 1-s stability near $2 \times 10^{-16}$. As seen from the Figure, the current level of the Doppler stabilized fiber links and optical synthesis with the OFC is sufficient to support the current best atomic clock comparisons. Currently, the 1-s instability of most inter-species clock comparisons is limited by the clock local oscillator. The development of cryogenic optical reference cavities \cite{Wiens2014,Matei2017} and differential spectroscopy of atomic clocks \cite{Kim2021} offer a path toward lower clock and measurement instability, respectively. As a result, optical synthesis with the OFC, as well as the dissemination of clock signals via optical fiber, will also need to improve to achieve better short term instability.

%%%%%%%%%%%%%%%%%%%%%%%%%%%%%%%%%%
%%%%%%%%%%%%%%%%%%%%%%%%%%%%%%%%%%
%%%%%%%%%%%%%%%%%%%%%%%%%%%%%%%%%%
\section{Stability and accuracy in optical-to-microwave synthesis}

Above we characterized the additive instability of the Er/Yb:glass laser OFC in optical synthesis and assessed its performance in the context of optical atomic clock comparisons. Here, we evaluate another important capability of OFCs, namely, their ability to coherently link the optical domain to the microwave domain. Briefly, when the OFC is phase locked to a high-stability cavity-stabilized optical reference, either free-running or locked to an atomic clock, the frequency stability of the optical reference is transferred to timing stability in the comb optical pulse train. Photodetection of the optical pulse train yields a pulse train in the electronic domain. Because the photodetector is only sensitive to the pulse envelope, the optical pulse carrier information is lost. As a result, the Fourier transform of the electronic pulse train only yields harmonics of the pulse repetition rate.

From the comb equation, to first order, the frequency of the comb repetition rate is proportional to $\nu_\text{ref}/N$, where $N$ is the mode number of the optical mode closest in frequency to the optical reference, $\nu_\text{ref}$. For this reason, the technique of optical-to-microwave conversion is called optical frequency division (OFD) \cite{Zhang2010,Haboucha2011,Fortier2013,Xie2017,Kalubovilage2020,Nakamura2020}. Not only is the optical frequency divided by $N$, but the frequency and phase noise power spectral density of the optical reference is also reduced by $N^2$. A 259 THz optical signal divided to 10 GHz, ideally, will yield a reduction in the optical phase noise by 88 dB ($20 \times \log[259 \text{ THz}/10 \text{ GHz}$]) and will preserve the optical fractional frequency stability in its conversion to the microwave domain. 

Optical frequency division permits the extraction of microwave timing signals from optical atomic references. The latter signals can be used for dissemination of optical atomic clock stability using traditional microwave networks, for their comparison and characterization against the $^{133}$Cs primary frequency standard, or for use as the local oscillator for advanced radar and communications systems. Below, we assess the additive instability, phase noise and frequency errors in optical-to-microwave synthesis.

%%%%%%%%%%%%%%%%%%%%%%%%%%%%%%%%%%%%%%%%%%
\subsection{Optical-to-microwave synthesis: 10 GHz phase noise}

The ability of an optical frequency comb to phase-coherently divide an optical signal to the microwave domain is crucial because it yields a straight-forward way to count optical clock oscillations with standard electronics. Additionally, optical-to-microwave synthesis provides a means to convert the low phase noise of atomic clock probe lasers to the microwave domain. The latter conversion helps realize high-spectral purity microwave signals, which have natural applications to next generation communications, sensing and ranging.

\begin{figure}
\centering 
\includegraphics[width=8.5cm]{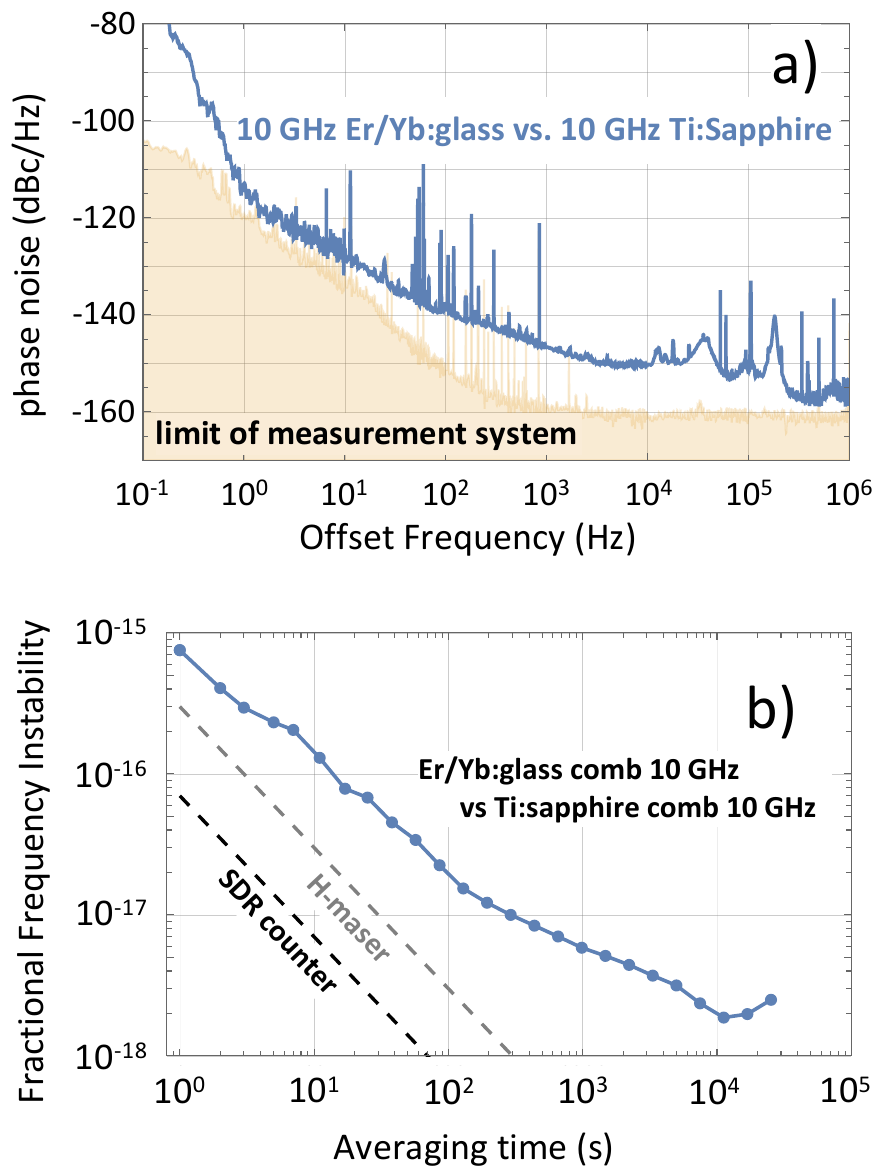}
\caption{Noise characterization in optical-to-microwave synthesis from 1157 nm to 10 GHz. a) Single sideband phase noise of the difference frequency, $\Delta f_{10\text{GHz}}$, between the 10 GHz generated via OFD by the Er/Yb:glass and Ti:sapphire combs. Also shown is the combined phase noise contributed by the 10 MHz H-maser reference (yellow) and the phase noise test set to the measurement at 10 GHz. b) Allan deviation of the difference frequency, $\Delta f_{10\text{GHz}}$, between the 10 GHz generated via OFD by the Er/Yb:glass and Ti:sapphire combs}
\label{OFD}
\end{figure}

In this section we assess the spectral purity of 10 GHz microwave signals derived via the 20$^\text{th}$ harmonic of the Er/Yb:glass laser repetition rate. As shown in Figure \ref{characterization-setup}, the 10 GHz generated by the Ti:sapphire comb is used as a reference signal for comparison \cite{Fortier2013}. To isolate the additive noise in optical-to-microwave synthesis, both OFCs were locked to the same 1157 nm optical reference. Additionally, both combs employed pulse interleavers that effectively multiply the comb repetition frequencies to 2 GHz prior to photodetection \cite{Haboucha2011}. Repetition rate multiplication improves the strength of the 10 GHz microwave harmonic by reducing the number of detected harmonics and helps to mitigate photodetector nonlinearities \cite{Taylor2011} that occur at high optical pulse energies. Detection nonlinearities and the microwave carrier strength were further improved by employing high-power, high-linearity modified uni-traveling carrier (MUTC) photodetectors \cite{Li2011}. The MUTC PD yields $\sim 12.5$ GHz bandwidth, 50 $\mu m$ active area diameter, and can be operated with a bias voltage as high as -21V \cite{Li2011,Li2010}. The photodetected 10 GHz signals from both combs were sent over 1 m of uncompensated microwave cable and compared using a double balanced mixer. The intermediate signal, $\Delta f_{10\text{GHz}}$ near 1 MHz, was measured using a Symmetricom 53132A phase noise analyzer that uses a 10 MHz H-maser signal as a reference. The results of this comparison are shown in Figure \ref{OFD} a), which characterize the additive noise of all components in Figure \ref{characterization-setup} except the 1157 nm optical cavity, which is a common source of noise for both frequency combs.

At 1 Hz offset from the 10 GHz carrier, we demonstrate a residual phase noise near -115 dBc/Hz. Frequency drift in the unstabilized microwave cables limits the phase noise for offset frequencies near and below 1 Hz. Phase noise at offset frequencies between 10 kHz to 200 kHz is the result of uncompensated noise in the $f_0$ and $f_\text{rep}$ locks of both optical frequency combs. Also shown is the additive noise of the 10 MHz H-maser reference to the 10 GHz phase noise measurement. We observe that the H-maser limits the 10 GHz comparison for offset frequencies below 10 Hz and contributes a high-frequency noise floor near -160 dBc/Hz. A lower measurement noise floor can be achieved by using a two channel phase noise measurement system with cross-correlation \cite{walls1992cross}. From past studies \cite{Fortier2013} that use cross-correlation, we expect a noise floor below -170 dBc/Hz. Assuming a thermal noise level of -170 dBc/Hz, we estimate a total integrated phase noise (from 1 Hz to 5 GHz) around 318 $\mu$rads, which indicates an RMS timing jitter on 10 GHz of around 7.2 femtoseconds.

%%%%%%%%%%%%%%%%%%%%%%%%%%%%%%%%%%%%%%%%%%
\subsection{Optical-to-microwave synthesis: additive instability}

To assess the additive instability for timescales longer than 1 s, we calculate the Allan deviation of the compared 10 GHz signals, $\Delta f_{10\text{GHz}}$. The results are depicted in Figure \ref{OFD} b). The instability represents an upper bound to the additive optical-to-microwave synthesis with the Er/Yb:glass laser OFC. This is because $\Delta f_{10\text{GHz}}$ carries phase information regarding the noise of the optical fiber links to the OFCs, the OFC optical-to-microwave synthesis, noise in photodetection and in the comparison and measurement of the 10 GHz signals. As seen in the Figure, the Allan deviation of $\Delta f_{10\text{GHz}}$ yields a 1-s additive instability $<1 \times 10^{-15}$, averaging down with a slope near $\tau^{-1}$.

%%%%%%%%%%%%%%%%%%%%%%%%%%%%%%%%%%%%%%%%%%
\subsection{Optical-to-microwave synthesis: additive synthesis errors}

To quantify the frequency errors added by optical-to-microwave synthesis, we generate and compare 10 GHz signals based on an absolute frequency reference. For the latter measurement, the Er/Yb:glass laser and Ti:Sapphire laser OFC were both phase stabilized to the same $^{171}$Yb optical atomic clock. Using the atomic reference, both combs generated electronic signals near 10 GHz. The Ti:Sapphire OFC generated a second signal near 1 GHz, $f_\text{rep,TiS}$, that was measured against the international ensemble of primary and secondary standards (PSFS), via the NIST maser ensemble \cite{NIST2021}. The difference between the two 10 GHz signals from the Ti:Sapphire and Er/Yb:glass laser OFCs, $\Delta f_{10\text{GHz}}$, was simultaneously counted on an SDR-based frequency counter, using a single H-maser in the NIST maser ensemble. As such, the Er/Yb:glass laser repetition rate can be evaluated as, $f_\text{rep,Er/Yb} = \left( 10 \times f_\text{rep,TiS} + \Delta f_{10\text{GHz}} \right)/20$. By simultaneously counting the offset frequencies and beat signals between the OFCs and their respective $^{171}$Yb local oscillator signals, the comb equation can be used to calculate the $^{171}$Yb optical frequency as measured by both OFCs.

\begin{figure}
\centering 
\includegraphics[width=8.5cm]{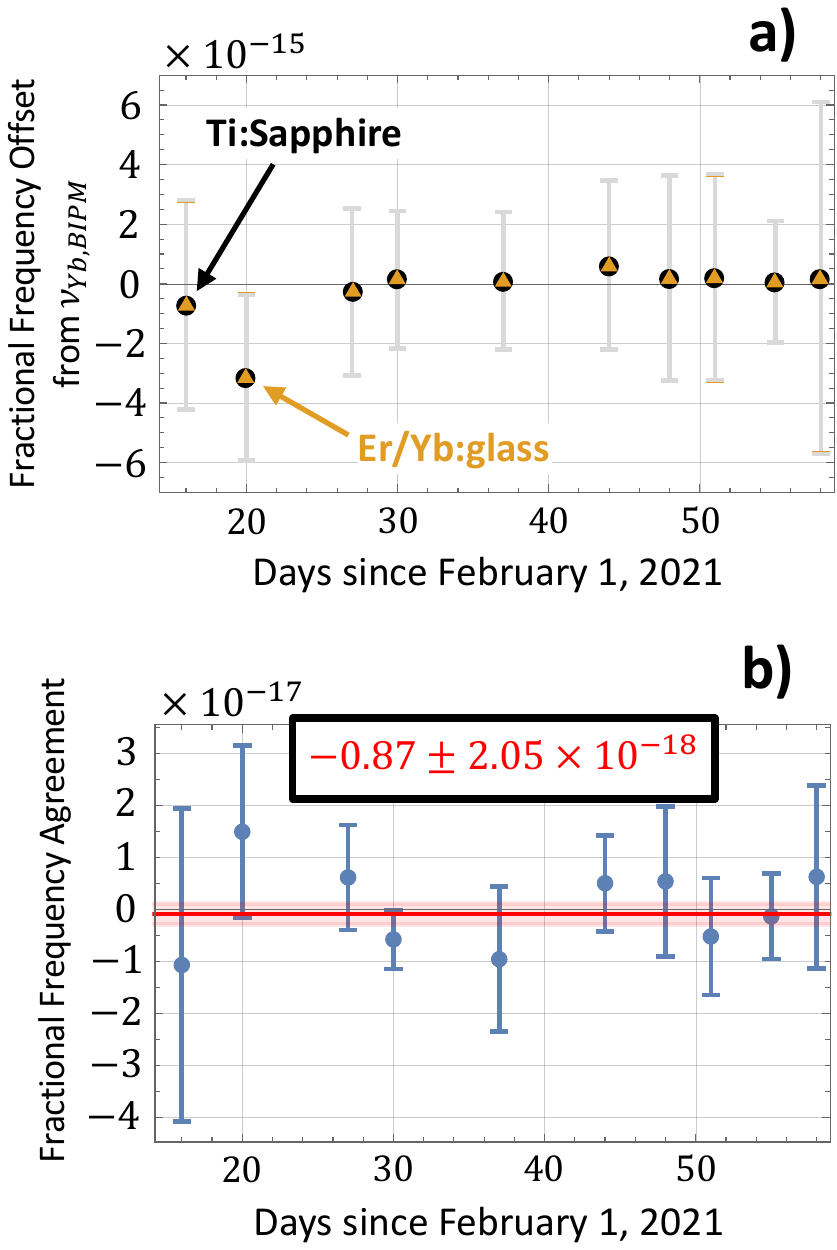}
\caption{a) Comparison in the evaluation of the absolute frequency of the $^{171}$Yb lattice clock as measured by the Er/Yb:glass laser OFC (yellow triangles) and Ti:sapphire OFC (black circles). Data in a) and b) were derived from unioned data only. Data in a) is offset from the BIPM recommended value for $^{171}$Yb of 518 295 836 590 863.63 Hz. The grey error bars were limited by the instability of the H-maser that was used for referencing the counted 1 GHz Ti:sapphire repetition rate signal.  b) Shows the fractional frequency difference between the black and yellow markers in a), demonstrating an agreement between the OFCs at the $10^{-18}$ level. The blue error bars in b) were limited by the instability of the H-maser that was used for referencing the counted $\Delta f_{10\text{GHz}}$ signal.}
\label{TO-absolute}
\end{figure}

Using this procedure we periodically measured the $^{171}$Yb optical atomic clock frequency vs PSFS using both frequency combs, Figure \ref{TO-absolute} a). In Figure \ref{TO-absolute} a), the error bars are determined by the daily statistical uncertainty, which is dominated by the noise of reference H-maser ($\sim 10^{-13}$ at 1-s averaging time), common to both OFCs. Figure \ref{TO-absolute} b) shows the difference between the Er/Yb:glass laser and Ti:Sapphire laser OFC measurements of the $^{171}$Yb optical atomic clock frequency, per day. The error bars in b) represent the statistical error in the measurement of $\Delta f_{10\text{GHz}}$, which is dominated by the additive noise of the SDR-based frequency counter. From Figure \ref{TO-absolute} b), we observe an agreement in the mean value of the daily measurement of the $^{171}$Yb optical clock, between both OFCs, at $-0.87 \pm 2.05 \times 10^{-18}$, limited by the measurement uncertainty. %The agreement indicates that the combined optical-to-microwave synthesis error of both combs is at the same level as the agreement in b).

%%%%%%%%%%%%%%%%%%%%%%%%%%%%%%%%%%
%%%%%%%%%%%%%%%%%%%%%%%%%%%%%%%%%%
%%%%%%%%%%%%%%%%%%%%%%%%%%%%%%%%%%

\section{Conclusion}

In summary, the Er/Yb:glass laser represents a versatile and robust system for laboratory-based, high performance metrology applications. The laser represents a competitive alternative to OFCs based on Er:fiber and Ti:sapphire modelocked lasers because it is simple to build and demonstrates self-starting modelocking due to the SESAM. Additionally, the low net laser cavity loss permits narrow intrinsic optical linewidths, despite the use of a 980 nm telecom diode laser for pumping. The latter permits high performance optical and microwave synthesis with moderate engineering of the laser feedback actuators and feedback loops. 

To convert the Er/Yb:glass laser to a frequency comb, we interfaced the laser with a home-built PM-Erbium-doped amplifier and a hybrid HNLF to enable generation of an octave-spanning optical spectrum. Once phase stabilized, we demonstrated the utility of the Er/Yb:glass laser OFC, and quantified its performance in terms of precision metrology applications. We have evaluated the additive noise in optical synthesis, which contributed an upper bound to the instability at 1-s near 10$^{-17}$. A lower bound to optical synthesis, below 10$^{-18}$, was found to be limited by uncompensated free-space optical paths in the interferometers used to reference the comb to high-stability optical frequency references. In 10 GHz optical-to-microwave synthesis, we observed an additive instability below 10$^{-15}$ for 1-s averaging times, with a residual single-sideband phase noise level below -115 dBc/Hz at a 1 Hz offset from the carrier. When phase-locked to a $^{171}$Yb optical atomic clock, we estimated that the Er/Yb:glass laser OFC contributed errors in optical-to-microwave frequency synthesis near the 10$^{-18}$ level, limited by the measurement statistics. 

Both high-precision optical-to-optical and optical-to-microwave synthesis are required for the implementation of an all-optical timescale \cite{Milner2019,Yao2019}, an important part of the infrastructure for redefinition of the SI second to optical atomic references. Additionally, optically derived timing signals will help realize timing signals that are at least two orders-of-magnitude more stable and accurate than current timescales based on microwave oscillators.

\section{Acknowledgements}
We thank J. A. Sherman for help with the SDR frequency counter and R. P. Mirin for the SESAM.

\section{Disclaimers}

Certain commercial equipment, instruments, or materials (or suppliers, or software, etc.) are identified in this paper to foster understanding. Such identification does not imply recommendation or endorsement by the National Institute of Standards and Technology, nor does it imply that the materials or equipment identified are necessarily the best available for the purpose.

\section{Data Availability}
The data that support the findings of this study are available from the corresponding author upon reasonable request.

\section*{References}

\bibliography{main.bib}

\section{Appendix / Supplementary}

%%%%%%%%%%%%%%%%%%%%%%%%%%%%%%%%%%
\subsubsection{cavity alignment procedure}

In this section we detail some of the steps taken to construct the modelocked Er/Yb:glass laser and specify characteristics that are useful in its optimization.

\textbf{Place optics:} We first secured the 980 nm pump fiber-to-free-space coupler. While the laser diode was running at low power we then secured components in the following order: (1) the $f=75$ mm lens, (2) the curved mirror (M2), (3) the crystal (at Brewster's angle, $\approx 57$ degrees) as seen in Figure \ref{setup}, and (4) the curved mirror (M2). Translation stages are used on all mirrors, except the curved mirror (M2) through which the 980 nm pump light is focused. In addition to a linear translation stage, the Er/Yb:glass gain medium is placed on a rotation stage to fine tune Brewster's angle. While actuation stages can reduce the alignment stability, they also permit higher efficiency in laser alignment. To reduce their impact to the optical alignment we employ a low beam height (4 cm), highly rigid optical mounts, and high-performance stainless actuator stages. A half-waveplate, after the 980 nm pump but before the focusing lens, may be necessary to tune the 980 nm polarization to minimize the Fresnel reflection off the Er/Yb:glass gain medium. 

For initial alignment, the green fluorescence from the gain medium is used to estimate the position of the lasing 1550 nm beam and its focal points. The latter technique is used to position the output coupler (OC) and SESAM, by centering the green fluorescence at their centers. Note that the SESAM is placed at the focus of the green fluorescence. To align the laser for CW operation, tip tilt on the OC and HR was used to overlap the fluorescent return beams from the two arms of the linear cavity; one between the Er/Yb:glass and the OC and the other from the Er/Yb:glass to the SESAM or HR.

Once the laser cavity is adequately aligned, the brightness of the green fluorescence from the crystal will dim noticeably as energy is transferred to the 1550 nm lasing light. When varying the pump current from threshold to higher power, the 1550 nm output power should also increase steadily. If the cavity alignment is sub-optimal, the effects of thermal lensing will be observed via thermally induced refractive index changes in the crystal. The latter manifest as sudden drops in output power as the pump power increases. 

After loosely positioning the output coupler, we then use a high-reflector in place of the SESAM to optimize the laser for continuous wave (CW) lasing. Using an HR in place of the SESAM for initial alignment also helps to avoid damage due to Q-switching. Once CW power is maximized, the high-reflector is replaced by the SESAM, using the fluorescence focus for positioning. Translation of the SESAM, via a linear translation stage, is used to induce and tune modelocked operation.

%\textbf{Optimize alignment:} Once the laser was lasing, we optimized the cavity alignment by maximizing the 1550 nm output power. Aside from using the tip/tilt action of the mirror mounts and the translation stages, there were several techniques we used to perfect the cavity alignment. We enumerate the most important below. 

%\item We turned down the pump power to find the lasing threshold. For optimal 980 nm to 1550 nm conversion, this threshold must be as low as possible. By tuning the cavity alignment while the pump power is low, we ensured that we were only optimizing for the most prominent cavity mode, and lowered the threshold to about 30 mW of pump power. Single-mode operation may also be checked with an optical spectrum analyzer.
%\item Once the crystal angle was close to Brewster's angle, we used the rotation stage to optimize further. We measured the power reflected from the crystal surface and minimized it by rotating the crystal while simultaneously optimizing the mirror angles to maximize power from the output coupler. Once optimized, the reflections from the crystal should total $< 1 \%$ of the power from the output coupler. 

\textbf{Modelock the laser:} Once the laser cavity was optimally aligned with the HR, we replaced the HR with the SESAM. With the 980 nm pump power near 400 mW, the correct SESAM position and tip-tilt angles were easily found by watching for the dimming of the green fluorescence. Once lasing, the threshold pump power was re-optimized by translating the SESAM through the 1550 nm focus. Once optimized, by increasing the pump power to > 250 mW the laser should generally begin modelocking. Finally, an optical spectrum analyzer was used to optimize the modelocked state by characterizing the CW breakthrough and spectral bandwidth as the SESAM is translated across the 1550 nm focus.

%Different modelocked states accessible as the SESAM was translated and also that the crystal location may be used to increase pulse bandwidth at the expense of output power. Additionally, it is also important to ensure that there is no CW breakthrough, which is easily seen with an RF spectrum analyzer as excess peaks that are not the 500 MHz repetition frequency or its harmonics.

% If in two-column mode, this environment will change to single-column format so that long equations can be displayed. 
%\begin{widetext}
%$$\mbox{put long equation here}$$
%\end{widetext}

% Figures should be put into the text as floats. 
% Use the graphics or graphicx packages (distributed with LaTeX2e).
% Put the label that you will use with \ref{} command in the braces of the \label{} command.
%
% \begin{figure}
% \end{figure}

%\begin{acknowledgments}
% Put your acknowledgments here.
%\end{acknowledgments}

\end{document}